# User Profile Feature-Based Approach to Address the Cold Start Problem in Collaborative Filtering for Personalized Movie Recommendation

Lasitha Uyangoda
Faculty of Information Technology
University of Moratuwa
Katubedda, Sri Lanka
lasitha.uyangoda@gmail.com

Supunmali Ahangama
Faculty of Information Technology
University of Moratuwa
Katubedda, Sri Lanka
supunmali@uom.lk

Tharindu Ranasinghe
Department of Research & Development
CodeGen International (pvt) Ltd.
Colombo 10, Sri Lanka
tharindu.10@cse.mrt.ac.lk

*Abstract*— A huge amount of user generated content related to movies is created with the popularization of web 2.0. With these continues exponential growth of data, there is an inevitable need for recommender systems as people find it difficult to make informed and timely decisions. Movie recommendation systems assist users to find the next interest or the best recommendation. In this proposed approach the authors apply the relationship of user feature-scores derived from user-item interaction via ratings to optimize the prediction algorithm's input parameters used in the recommender system to improve the accuracy of predictions when there are less past user records. This addresses a major drawback in collaborative filtering, the cold start problem by showing an improvement of 8.4% compared to the base collaborative filtering algorithm. The user-feature generation and evaluation of the system is carried out using the 'MovieLens 100k dataset'. The proposed system can be generalized to other domains as well.

## I. Introduction

With the advancement of the technology, the types of items consumed, and the way they are consumed have changed. With the limitless exposure to information makes a consumable that was previously unknown, to be a sort after item within few seconds. For example, a previously unknown music video like "Gangnam Style" [1] from an unpopular artist can be made popular within few seconds. Even though there are many options to consider, the chances of going through all of them in a single lifetime is impossible. Thus, there is a real necessity of selecting the best option from the possible to-do list. To discover the best option, within a limited life span available, consumers tend to refer to reviews made by other consumers of the product or service. However, considering the information overload, it is not possible to make timely decisions. Thus, to overcome the information overload problem and to assist effective decision-making process, recommender systems are developed.

In this paper, recommendation of movies is considered as there are millions of movies available to select and each movie in average has a length around 90 minutes. Practically, it is impossible to exceed the watch-count to thousand where most people tend to watch less than thousand movies during their life span. Therefore, people tend to depend more on recommender systems to identify best movies to watch.

The intention of this study is to propose a more personalized recommender system to recommend movies to the users. The system is modeled using the 'MovieLens_100k' data set which has a rich amount of details regarding users, items and user-item interaction. In this study, with the identification of the possibility to create user features using an existing general 'user item interaction' data records, it was possible to improve the accuracy to a significant level when dealing with the cold start problem.

In the subsequent sections of paper, background details of recommender systems, recommendation techniques, descriptive analysis of the dataset, proposed algorithm, evaluation results and further works will be discussed in detail.

## II. Background

Users and items are the two types of things that need to exist to give recommendations. A recommender system to work, users should provide feedback about items. There are two types of feedbacks [2];

- Implicit: the feedback that is provided unintentionally, for an instance; buying an item through E-bay.

- Explicit: the feedback that is provided intentionally, for an instance; rating/reviewing a movie that you have watched on IMDB.

A recommender system has to identify above feedbacks and predict or estimate the ratings of items that a particular user has not yet interacted with. Therefore, a recommender system is more like a function that matches a user-item pair to a rating which is the prediction itself of the recommender system. Then can be used to evaluate the accuracy of the recommender system considering whether the item will be actually rated by the user.

### A. Cold-Start Problem

The cold start problem, also known as the bootstrapping problem, when there are not enough, or no previous history related to a user's interaction with items within the recommendation system. In a situation like this, the ability of giving meaningful recommendations to the user is no possible [3].

## III. Existing Approaches

There are various types of recommender systems used to satisfy today's recommendation requirement. Also, the number of recommendation techniques used in the system depends on the system requirement. This explains the importance of understanding the features and potentials of different recommendation techniques. Fig. 1 shows the anatomy of different recommendation filtering techniques [4].

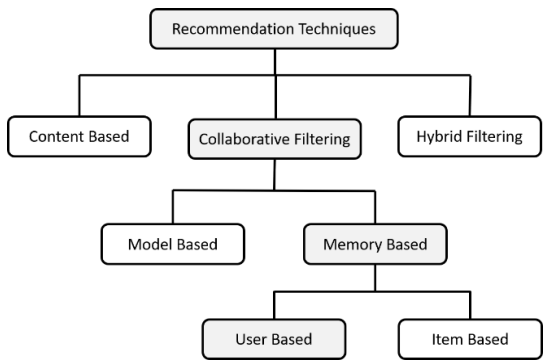

Fig. 1. Different types of recommendation techniques.

*A. Considered Recommendation Technique*

The used approach can be varied depending on the requirement and the intention of the research. In this research memory-based collaborative filtering was used.

*1) Collaborative Filtering (CF)*

Collaborative filtering is a domain-independent prediction technique for content that cannot be easily and adequately described using metadata of movies and music. Collaborative filtering techniques work by building a database (user-item matrix) of preferences for items by users. It then matches users with relevant interest and preferences by calculating similarities between their profiles to make recommendations [4]. CF can be further divided into two categories: memory-based and model-based [5] [6].

*2) CF-Memory Based*

The items that were already rated by the user play an important role in searching for a neighbor that shares similar types of appreciation [7] [8]. Once a neighbor of a user is found, different algorithms can be used to combine the preferences of neighbors to generate recommendations.

*B. Similar Works*

In order to address with the Cold-Start Problem, many researches have been conducted covering various aspects. Such as;

The content enhanced approach [13] identified by Dongting Sun, Cong Li and Zhigang Luo in which the novel user is pre-clustered based on his/her demographical information and given recommendations then combine the recommendation with the original collaborative filtering algorithm, once user creates enough records.

According to Peng Yi, Chen Yang, Xiaoming Zhou, Chen Li. Their approach [14] by bridging the gap between movies and their labels used to identify the content has shown a significant improvement in Cold-Start Problem further they propose a methodology to optimize the similarity measurement by computing the similarity among directors(movie) and actors(movie).

By introducing the Association Cluster Filtering (ACF) algorithm [15], to create cluster models based on user-rating matrix, Chuangguang Huang and Jian Yin have proposed a solution for the users with less rating history by putting them into clusters and have assumed that each cluster precisely represents its users.

According to Simon Fong, Yvonne Ho, Yang Hang when there is a large population of variables to be considered to make recommendations, Genetic Algorithm (GA) [16] search function optimizes the recommendation results which also reduce the Cold-Start Problem. Further the process of coding the variables into GA chromosomes in various mods is explained.

IV. DATASET ANALYSIS

*A. Overview*

Before adapting the data set to build user feature profiles, it is necessary to analyze the data set to get an insight to it to improve the algorithm. In this study, MovieLens_100k data set [9] collected by the GroupLens is used.

- User Data - Basic demographic details of the user.
- Item Data – Basic details of the movie and genre (19 different genres) wise content information.
- User Item Interaction Data – User id, Item id and the rating given to each item(movie).

*B. Descriptive Analysis*

*1) Age vs Users*

Fig. 2 shows the age distribution in the dataset. The dataset is dense round age 30. As there is no uniform distribution of age, it would difficult to identify the influence of age.

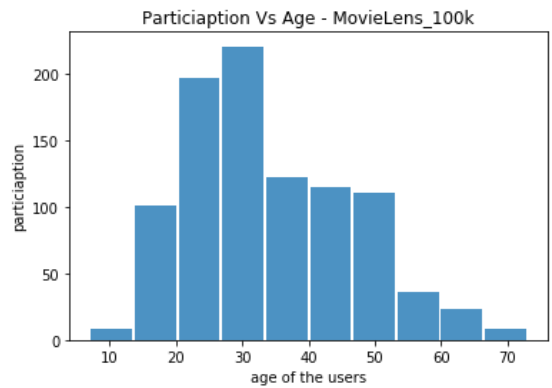

Fig. 2. No of participants based on age.

*2) Number of users based on their Occupation*

Fig. 3 shows the number of participants based on their occupation. The participation of students is comparatively very higher than all the other occupations. Therefore, the occupation wise user distribution in the dataset is highly uneven. This result is in line with the age distribution result (Fig. 2).

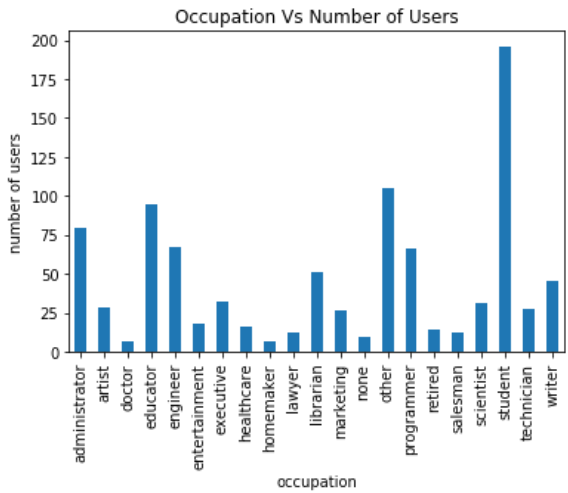

Fig. 3. No of participants based on occupation.

*3) All genre-encounters based on Occupation*

With the insights gained from Fig. 2 and Fig. 3, it was expected to find a noticeable genre encounter level deviation from the students-users which means the expectation was to find out the most popular genre of each occupation. Genre encounter level number describes how much each genre has occurred in all the single watches done by all the users cumulatively. When all the users are categorized according to their occupation to get the total genre encounter levels for each occupation. Rather surprisingly, the actual result was completely different. As per Fig. 4, despite the genre and its encounter level for each occupation, it could be observed that the pattern of genre encounter level for each occupation is nearly similar. Therefore, based on that we can assume that the ratio of genre levels in a single occupation does not change over the consumption of movies. According to that comparing two users from different occupations and different age levels with different encounter levels is effective.

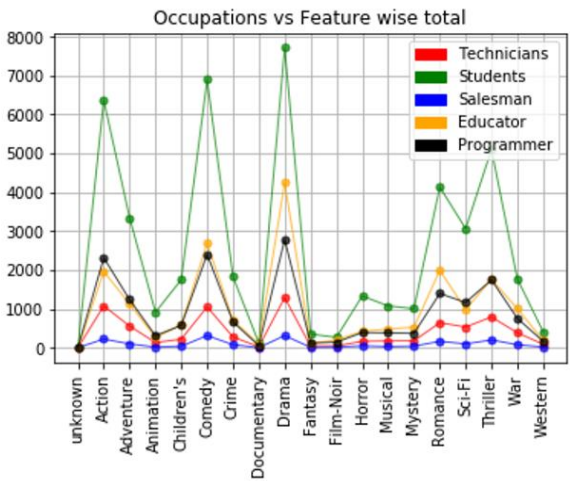

Fig. 4. Encounter level based on occupation.

## V. PROPOSED MODEL

### A. Overview

The main goal of this study is to personalize each user preferences based on the feature scores generated out of the previous ratings given to the items. Since all the users are having their independent ratings on the items they consumed, the pattern they generate is identical to each user. Thus, identical data from the rating history was extracted and a feature score profile for each user was created. Those feature scores were used to identify similar users. This model contains three major modules;

- Creating user feature profiles based on past records.
- Calculate similarity between profiles based on feature scores.
- Generate recommendations based on created similarity matrix.

The process model of this study is shown below in Fig. 5. It shows the basic flow of main inputs, outputs and processes involved.

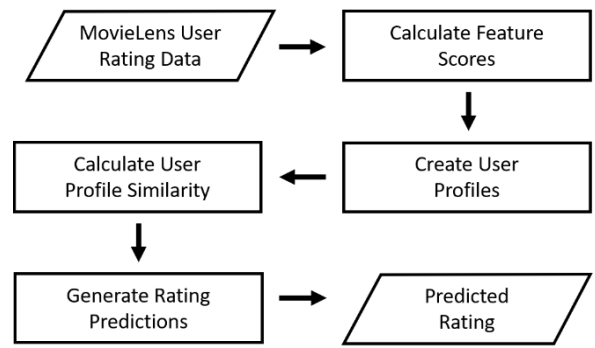

Fig. 5. Process model of the study.

### B. Overview of the process model

This section explains the flow of the prosses model.

*1) Feature score generation from user-item interaction data*

The idea of generating common features for users is to improve the accuracy of similarity measurement where users become more identical from one another. To achieve this, we consider the list of existing genres (19) as features. For a given user, each movie and the rating he/she has given to that movie is distributed among all the genres that movie has.

Ex: Movie 'A' has Action, Romance and Thriller. If the user given rating is 3/5, that is distributed as Action = 3/5, Romance = 3/5 and Thriller = 3/5

From the user's point of view, it can be considered that each genre/feature score is corresponding to genre/feature wise entertainment received by the user from that movie.

Through this approach a personalized user profile is created based on their past interactions with the movies.

Feature wise number of encounter will give an idea about how much movies he/she has watched contain that feature.

Ex. movie A has (Action, Comedy, Drama, Romance) and movie B has (Action, Thriller, Romance). If user X watched those two movies, his/her encounter numbers will be (Action-2, Comedy-1, Drama-1, Romance – 2, Thriller-1)

For each movie and its rating, feature scores can be generated for all the genres that movie is about (ex: Romance, Drama, etc.). In the movie(item) data set, relevant genres are indicated using binary indication (0 – does not contain this genre, 1- contains this genre) which is used as the features. Following equation was developed to calculate above explained feature score;

feature-f, current feature score-c and the rating given-R, number of encounters – n, and the updated feature score, *f.s*:

$$f.s = \frac{(f \times R) + (c \times n)}{(n + f)} \quad (1)$$

If feature score, *f.s* is very low and the number of encounters, n is very high, then the user probably does not like that feature (genre).

*2) User Profiles*

Profiles contain basic demographic details about the user, feature scores (where 19 genres in the dataset are considered as the features) and the number of encounters of each feature.

E.g. User (0234) has watched 34 movies that contained the feature (genre) 'drama'. Thus, the current 'drama-feature' score is 3.65/5.

TABLE I.  FEATURE SCORE OF THE USER

| user id | feature score (Drama) | No of encounters (Drama) |
|---|---|---|
| 0234 | 3.65/5 | 34 |

When the user watches a new movie that has the genre 'drama' and if he/she rates the movie with a rating of 4/5 then the new feature score would be 3.66/5.

TABLE II.  UPDATED FEATURE SCORE OF THE USER

| user id | feature score (Drama) | No of encounters (Drama) |
|---|---|---|
| 0234 | 3.66/5 | 35 |

*C. Similarity Measurement*

Pearson Similarity is used as the similarly measure between user profile features. It is well established similarity calculation method and is being widely used in various context [10].

## VI. EVALUATION

Quality of a recommendation algorithms can be examined with the use of different evaluation criteria for accuracy or coverage of the algorithm. Type of metrics used to evaluate are dependent on the filtering technique used in the algorithm  Accuracy of the recommendation is taken as the fraction of the correct recommendations  from the total recommendations made.

Statistical accuracy metrics evaluate the accuracy of a filtering technique by comparing the predicted ratings directly with the actual user rating. Mean Absolute Error (MAE) [11], Root Mean Square Error (RMSE) and Correlation are usually used as statistical accuracy metrics.

RMSE is the most popular and commonly used method to evaluate the recommender algorithms [12].

The results derived from the evaluation are shown in the Table III. Algorithms mentioned in the table are;

- Base Collaborative Filtering Algorithm (BCF) – Basic implementation of Collaborative Algorithm using weighted average method.

- Modified Collaborative Filtering Algorithm (MCF) – Similarity input of the users are measured between user profile features.

- MCF only with maximum number of five records - user profile features are generated only with a maximum of five random records.

BCF algorithm kept increasing the error with the reduction of records fed to find the similarities where MCF kept decreasing the error for less records and scored its best at up to maximum of five records as shown in the Table III.

TABLE III.  EVALUATION RESULTS

| Algorithm | RMSE Value |
|---|---|
| BCF | 1.1874 |
| MCF | 1.1823 |
| BCF using five records | 1.2737 |
| MCF using five records | 1.1672 |

MCF for five random records has the lowest RMSE value. Overall performance of the algorithms has improved by 4.3% and only with five random records, it has further improved by 8.4%. This improvement is obtained mainly because of MCF uses the user profile feature similarities of the users to identify the similar users where the BCF uses the rating history of the users where two feature level similar users (according to MCF similarity) can be ignored because they do not have a common rating history.

According to the results shown in TABLE III, performance of the MCF is ahead of BFC with lower number of records. Therefore, MCF can be used in recommendation systems where the novel user density is higher compared to mature users. Most of the similar approaches taken to solve the Cold-Start Problem in this domain [13,14,15] uses additional input parameters or complex calculations where the simplicity of the algorithm is compromised. In this particular finding in MCF, the simplicity of the algorithm is well maintained. Therefore, when it comes to the improvement of an algorithm, it should also be about simplicity which enhances the efficiency and performance.

Modern recommender systems are dealing with a huge number of user data and item data obviously consume a lot of processing power to manipulate. According to Occam's razor [17] which is also applicable in Modern Machine Learning, explains the necessity of using the least complex to deploy and easiest to understand algorithm.  Simple the algorithm produces faster and efficient results which are important features in a modern system [18]. Therefore, the

MCF algorithms explained in this paper is well ahead of other similar approaches to address the Cold-Start Problem.

## VII. DISCUSSION

With the results acquired from MCF, the accuracy improvement of 8.36% only with maximum of five user record is a significant finding in addressing the cold-start problem which is a major drawback in Collaborative Filtering. Compared to other related works [13,14,15] discussed above in the Existing Approaches section, MCF performs on par with related solutions and yet be the simplest algorithms to both understand and implement. the results of the evaluation, the improving accuracy under less records and delivering the best accuracy with less records rather than with more records happens to be the best finding in this research which requires further studies and improvements.

## VIII. FURTHER WORKS

Using the same profile feature generating approach, creating movie profiles that involves Content Based Filtering approach will a great advantage to improve the existing recommendation engine to a hybrid model may further improve the recommendations. Moreover, finding the influence of the number of encounters of each features is another important factor to be considered in future studies. To utilize the maximum advantage of this simple but yet powerful algorithm, certain optimizations should be conducted and should be tuned for the optimum performance. Finally, to improve the generalizability, the model can be tested with different datasets in different domains.

## IX. CONCLUTION

With this novel technique of adapting user-item interaction history into user feature scores, opens many doors to improve existing collaborative algorithm without sticking to user-item rating data. Simple adaptation of the exiting user records data to create user profile features consumes less processing power which make this algorithm more comfortable to introduce into existing recommendation systems. With the increasing demand for recommendations over almost everything on the internet and every digital platform, this finding will stand out by giving a better accuracy under minimal processing power which guides the recommendation algorithm field into a greener future.

## REFERENCES

[1] en.wikipedia.org/Gangnam_Style

[2] Hu, Y., Koren, Y., and Volinsky, C. Collaborative Filtering for Implicit Feedback Datasets. 2008 Eighth IEEE International Conference on Data Mining, (2008), 263-272.

[3] S. T. Park e W. Chu, "Pairwise preference regression for cold-start recommendation," em RECSYS '09: Proceedings of the third ACM Conference on Recommender Systems, New York, 2009.

[4] J. L. Herlocker, J. A. Konstan, L. G. Terveen, and J. T. Riedl, "Evaluating collaborative filtering recommender systems," ACM Trans. Inf. Syst., vol. 22, no. 1, pp. 5–53, Jan. 2004.

[5] ] J. Bobadilla, F. Ortega, A. Hernando, and A. Gutiérrez, "Recommender systems survey," Knowl.-Based Syst., vol. 46, pp. 109–132, Jul. 2013.

[6] ] J. S. B. David and H. C. Kadie, "Empirical Analysis of Predictive Algorithms for Collaborative Filtering" p. 10.

[7] F. O. Isinkaye, Y. O. Folajimi, and B. A. Ojokoh, "Recommendation systems: Principles, methods and evaluation," Egypt. Inform. J., vol. 16, no. 3, pp. 261–273, Nov. 2015.

[8] X. Zhu, H. Ye and S. Gong, "A personalized recommendation system combining case-based reasoning and user-based collaborative filtering," 2009 Chinese Control and Decision Conference, Guilin, 2009, pp. 4026-4028.

[9] F. Maxwell Harper and Joseph A. Konstan. 2015. The MovieLens Datasets: History and Context. ACM Transactions on Interactive Intelligent Systems (TiiS) 5, 4, Article 19 (December 2015), 19 pages.

[10] Hyung Jun Ahn. A new similarity measure for collaborative filtering to alleviate the new user cold-starting problem. Information Sciences 178 (2008) 37–51.

[11] Cotter P, Smyth B. PTV: Intelligent personalized TV guides. In: Twelfth conference on innovative applications of artificial intelligence; 2000. p. 957–64.

[12] K. Goldberg, T. Roeder, D. Gupta, C. Perkins Eigentaste: a constant time collaborative filtering algorithm Inform Retrieval J, 4 (2) (2001), pp. 133-151

[13] Dongting Sun, Cong Li and Zhigang Luo, "A content-enhanced approach for cold-start problem in collaborative filtering," 2011 2nd International Conference on Artificial Intelligence, Management Science and Electronic Commerce (AIMSEC), Dengleng, 2011, pp. 4501-4504.

[14] P. Yi, C. Yang, X. Zhou and C. Li, "A movie cold-start recommendation method optimized similarity measure," 2016 16th International Symposium on Communications and Information Technologies (ISCIT), Qingdao, 2016, pp. 231-234.

[15] C. Huang and J. Yin, "Effective association clusters filtering to cold-start recommendations," 2010 Seventh International Conference on Fuzzy Systems and Knowledge Discovery, Yantai, 2010, pp. 2461-2464.

[16] S. Fong, Y. Ho and Y. Hang, "Using Genetic Algorithm for Hybrid Modes of Collaborative Filtering in Online Recommenders," 2008 Eighth International Conference on Hybrid Intelligent Systems, Barcelona, 2008, pp. 174-179.

[17] en.wikipedia.org/Occam's_razor

[18] Y. M. Yufik and T. B. Sheridan, "Swiss army knife and Ockham's razor: modeling and facilitating operator's comprehension in complex dynamic tasks," in IEEE Transactions on Systems, Man, and Cybernetics - Part A: Systems and Humans, vol. 32, no. 2, pp. 185-199, March 2002.